\documentclass[conference]{IEEEtran}
\usepackage{cite}
\usepackage{enumitem}
\usepackage{ragged2e}
\usepackage{tabularx}
\usepackage{array}
\usepackage{pgf}
\usepackage{amsmath,amssymb,amsfonts}
\usepackage{algorithmic}
\usepackage{graphicx}
\usepackage{textcomp}
\usepackage{float}
\usepackage{xcolor}
\usepackage[section]{placeins}
\usepackage{braket}
\usepackage[hidelinks]{hyperref}

\def\BibTeX{{\rm B\kern-.05em{\sc i\kern-.025em b}\kern-.08em
    T\kern-.1667em\lower.7ex\hbox{E}\kern-.125emX}}
\newcolumntype{Y}{>{\centering\arraybackslash}X}

\begin{document}
    \title{Quantum Opacity, Classical Clarity: A Hybrid Approach to Quantum Circuit Obfuscation}
    
    \author{\IEEEauthorblockN{Amal Raj}
    \IEEEauthorblockA{\textit{Infocomm Technology} \\
    \textit{Singapore Institute of Technology}\\
    Singapore \\
    amal.raj@singaporetech.edu.sg}
    \and
    \IEEEauthorblockN{Vivek Balachandran}
    \IEEEauthorblockA{\textit{Infocomm Technology} \\
    \textit{Singapore Institute of Technology}\\
    Singapore \\
    vivek.b@singaporetech.edu.sg}}

    \maketitle
    
    \begin{abstract}
        Quantum computing leverages quantum mechanics to achieve computational advantages over classical hardware, but the use of third-party quantum compilers in the Noisy Intermediate-Scale Quantum (NISQ) era introduces risks of intellectual property (IP) exposure. We address this by proposing a novel obfuscation technique that protects proprietary quantum circuits by inserting additional quantum gates prior to compilation. These gates corrupt the measurement outcomes, which are later corrected through a lightweight classical post-processing step based on the inserted gate structure. Unlike prior methods that rely on complex quantum reversals, barriers, or physical-to-virtual qubit mapping, our approach achieves obfuscation using compiler-agnostic classical correction. We evaluate the technique across five benchmark quantum algorithms—--Shor's, QAOA, Bernstein-Vazirani, Grover’s, and HHL—--using IBM’s Qiskit framework. The results demonstrate high Total Variation Distance (above 0.5) and consistently negative Degree of Functional Corruption (DFC), confirming both statistical and functional obfuscation. This shows that our method is a practical and effective solution for the security of quantum circuit designs in untrusted compilation flows.
    \end{abstract}
    
    
    \section{Introduction}
        Quantum computing is an emerging field that utilizes the principles of quantum mechanics to achieve exponential speed-ups for certain tasks compared to even the most powerful classical computers \cite{ibm}. By exploiting quantum properties such as superposition and entanglement, quantum bits---or qubits---can exist in multiple states simultaneously, allowing quantum computers to tackle problems that are computationally hard for classical systems, such as factoring large numbers \cite{shor}, simulating molecular interactions \cite{molecular-interaction}, or optimizing complex systems \cite{optimization}. This potential has led to extensive research in the field, with leading institutions and companies driving quantum computing forward. Developments such as Google's Willow chip \cite{google} and Microsoft's pursuit of topological qubits based on Majorana fermions \cite{microsoft} are all indicative of the growing interests in making quantum computing practical. 

        However, the current era of quantum computing, called the Noisy Intermediate-Scale Quantum (NISQ) \cite{nisq} era, is marked by significant technical challenges. NISQ devices operate with limited numbers of qubits, and are prone to noise and decoherence, meaning their quantum states degrade rapidly over time \cite{noise-survey}. This noise causes errors in calculations, limiting the complexity and depth of algorithms that can be run securely. To bridge this gap, developers often depend on third-party quantum compilers to optimize the circuit. Compilers are tools that translate high-level quantum algorithms into machine-readable instructions tailored to a specific hardware. Several companies have developed compilers catering to their hardware platforms, such as IBM's Qiskit \cite{qiskit}, Quantinuum's t$\ket{\text{ket}}$\cite{tket}, and Rigetti's Quilc \cite{quilc}, among others. 
        
        This reliance on external tools necessitates the need to safeguard sensitive quantum code. When untrusted compilers receive proprietary algorithms, the potential exists for valuable IP to be reverse-engineered, stolen, or abused, as quantum computing becomes increasingly of commercial interest. To counter these risks, researchers and developers have been investigating obfuscation mechanisms, which seek to hide code structure and purpose without affecting its functionality. Addressing these challenges is crucial as quantum computing advances from theoretical promise to practical reality.

        \subsection{Proposed Idea}
\begin{figure}[htbp!]
                \centering
                \includegraphics[width=1\linewidth]{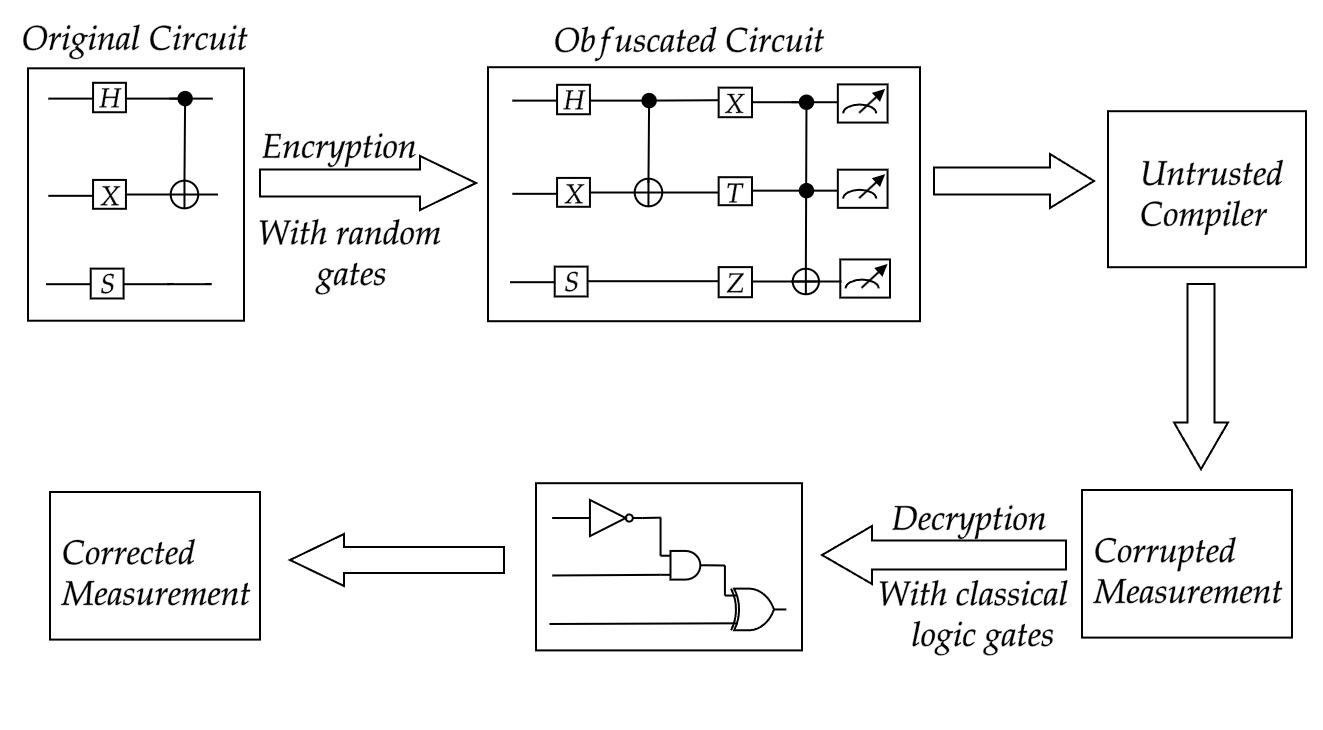}
                \caption{Workflow of our proposed idea}
                \label{fig:workflow}
            \end{figure}

            We propose a novel quantum obfuscation technique that simplifies the deobfuscation process by leveraging classical post-processing. The technique involves inserting additional quantum gates into some or all qubits of the original quantum circuit prior to compilation by an untrusted third-party compiler. These gates can be selected randomly or based on the gates in the original circuit, with the only constraint being the exclusion of Hadamard gates or their controlled versions, as they would introduce superposition which is difficult to correct classically. Once the circuit is compiled and executed, the resulting classical measurement outcomes are significantly corrupted due to the presence of the inserted gates. These outputs are then passed through a classical correction circuit, constructed based on knowledge of the inserted ``encryptor'' gates, to recover the actual measurement results --- effectively reversing the obfuscation classically. A diagrammatic representation of the same is shown in Fig. \ref{fig:workflow}.
            
            Existing obfuscation techniques for quantum circuits employ various strategies. One approach involves inserting a random quantum circuit into the original design. Its inverse is then compiled using another quantum compiler and inserted at the appropriate location after the original circuit has been compiled \cite{randomized-reversible}. This method implicitly assumes that the obfuscator or developer has access to a personal quantum compiler or quantum computer, which may not always be practical or scalable. Other techniques include inserting dummy gates that are subsequently removed during optimization \cite{dummy-gates}, or applying logic locking methods to restrict functionality without the correct key \cite{qll,e-loc}. These methods often require barriers to mark insertion points or knowledge of the physical-to-virtual (p2v) qubit mappings, making deobfuscation non-trivial and error-prone. By utilizing classical post-processing for decryption, our approach eliminates the need for such indicators or intricate quantum reversals. By avoiding the need to track insertion locations and utilizing classical post-processing for decryption, we reduce overhead and improve practicality when working with untrusted compilers.
            
        \subsection{Organization of the Paper}
            The rest of this paper is structured as follows. Section \ref{sec:background} describes background on quantum computing, covering qubits, quantum gates, compilation, and measurement operations. Section \ref{sec:related-works} is a survey of current works on obfuscating quantum circuits, citing current techniques and their shortcomings. Section \ref{sec:metrics} is the introduction of the evaluation metrics, Total Variation Distance (TVD) and Degree of Functional Corruption (DFC), used to quantify the effectiveness of obfuscation. Section \ref{sec:proposed-technique} describes our suggested obfuscation technique, its methodology, and an example application of the QAOA circuit. Section \ref{sec:evaluation} demonstrates the simulation results on a range of quantum algorithms, which verifies the effectiveness of the approach. Section \ref{sec:limitations} describes the shortcomings of our research and proposes directions for future work. Section \ref{sec:conclusion} summarizes the paper, including key contributions and implications.
            
    \section{Background}\label{sec:background}
        \subsection{Basics of Quantum Computing}
            \subsubsection{Qubits}
                Quantum bits or qubits are the fundamental units of quantum computing, analogous to bits in classical computing. While a classical bit exists only in one of the two states, 0 or 1, a qubit can exist in either of its basis states, $\ket{0}$ or $\ket{1}$, or as a linear combination of these states, known as superposition. Mathematically, the state of a qubit $\ket{\psi}$ is expressed as:
                
                \begin{equation}
                    \ket{\psi} = \alpha \ket{0} + \beta \ket{1},
                \end{equation}
                
                where $\alpha$ and $\beta$ are complex numbers satisfying the normalization condition $|\alpha|^2 + |\beta|^2 = 1$. Here, $|\alpha|^2$ and $|\beta|^2$ represent the probabilities of measuring the qubit in states $\ket{0}$ and $\ket{1}$, respectively \cite{quantum-computing}.
    
            \subsubsection{Quantum Gates}
                Quantum gates are the building blocks of quantum circuits, analogous to logic gates in classical computing. They are represented by unitary matrices, which ensure reversibility—a key property of quantum computation. A matrix $U$ is said to be unitary if it satisfies the condition $UU^{\dagger} = I$, where $U^{\dagger}$ represents the conjugate transpose of the matrix $U$, and $I$ is the identity matrix. Quantum gates operate on one or more qubits to perform transformations such as bit flips, phase shifts, or entanglement creation. Common gates include:
                
                \begin{enumerate}[label=\roman*.]
                    \item Pauli-X gate: Flips the state of a qubit from $\ket{0}$ to $\ket{1}$ and vice-versa.
                        \begin{equation*}
                            X = \begin{bmatrix} 0 & 1 \\ 1 & 0 \end{bmatrix}
                        \end{equation*}

                    \item Pauli-Y gate: Combines a bit flip and a phase shift, transforming $\ket{0}$ to $i\ket{1}$ and $\ket{1}$ to $-i\ket{0}$.
                        \begin{equation*}
                            Y = \begin{bmatrix} 0 & -i \\ i & 0 \end{bmatrix}
                        \end{equation*}
                        
                    \item Pauli-Z gate: Introduces a phase change if the qubit is in state $\ket{1}$, and leaves $\ket{0}$ unaffected.
                        \begin{equation*}
                            Z = \begin{bmatrix} 1 & 0 \\ 0 & -1 \end{bmatrix}
                        \end{equation*}

                    \item Hadamard (H) gate: Transforms either basis states to a superposition with equal probabilities of $\ket{0}$ and $\ket{1}$.
                        \begin{equation*}
                            H = \frac{1}{\sqrt{2}} \begin{bmatrix} 1 & 1 \\ 1 & -1 \end{bmatrix}
                        \end{equation*}

                    \item Controlled-NOT (CNOT) gate: A two-qubit gate that flips the target qubit’s state if the control qubit is $\ket{1}$, entangling the qubits if the control is in superposition.
                        \begin{equation*}
                            CNOT = \begin{bmatrix} 1 & 0 & 0 & 0 \\ 0 & 1 & 0 & 0 \\ 0 & 0 & 0 & 1 \\ 0 & 0 & 1 & 0 \end{bmatrix}
                        \end{equation*}
                \end{enumerate}
                

            \subsubsection{Compilation Process}
                Quantum compilation is essentially about translating an arbitrary high-level quantum algorithm into an implementation which can be run over an underlying quantum hardware platform. In an ideal quantum computing paradigm, all qubits can be connected to each other, and in an ideal setup, any quantum gate can be composed in front of any qubit or any combination of qubits. Realistic quantum hardware, especially in the NISQ era, places strong constraints on this ideal notion: they have only a few native gates (called basis gates) and restricted qubit interconnectivity, given by an underlying coupling map \cite{compilation}. The coupling map is often modeled as a graph, whose vertices represent physical qubits and whose edges represent pairs of qubits that can be directly connected using two-qubit gates. Compilation fills this gap between theoretical algorithms and feasible hardware by using the following well-defined intermediate steps:
                
                \begin{enumerate}[label=\roman*.]
                    \item Gate Synthesis: Complex gates (e.g., controlled-Z) are broken into basis gates like CNOT and single-qubit rotations.
                    \item Qubit Mapping: Logical qubits in the algorithm are assigned to physical qubits on the hardware, adhering to the coupling map. This physical-to-virtual (p2v) mapping can be complex, as the initial assignment may not align with connectivity requirements. If a two-qubit gate targets non-adjacent qubits, SWAP gates—each costing three CNOTs—are inserted to reposition qubit states, increasing circuit depth and error potential.
                    \item Optimization: Redundant gates are removed, and the circuit is refined to reduce errors.
                    \item Native Conversion: The circuit is translated into the device’s gate set (e.g., \{RZ, X, CZ\}).
                \end{enumerate}
            
            \subsubsection{Measurement Operation}
                Measurement collapses a qubit’s superposition into a classical state, either 0 or 1, with probabilities determined by its amplitudes. For a qubit $\ket{\psi} = \alpha \ket{0} + \beta \ket{1}$, the probability of measuring $\ket{0}$ is $|\alpha|^2$, and $\ket{1}$ is $|\beta|^2$. Post-measurement, the qubit loses its quantum properties, becoming a classical bit. For instance, measuring $\ket{\psi} = \frac{1}{\sqrt{2}} \ket{0} + \frac{1}{\sqrt{2}} \ket{1}$ yields 0 or 1 with equal probability, and the system collapses to the observed state.
    
    \section{Related Works}\label{sec:related-works}
        In the past decade, many different obfuscation techniques have been developed for quantum circuits. Our threat model aligns very closely with that of the work discussed in \cite{dummy-gates}, which proposed the inclusion of a dummy CNOT gate for optimal corruption of the functionality of a quantum circuit. Barriers are placed around the dummy gate to facilitate post-compilation removal, and around the other gates to mask its presence. Although efficient, this approach loses circuit optimization since the compilers cannot optimize across barriers, and the barriers may be seen as a sign of the presence of the dummy gate, thereby compromising security.

        Das and Ghosh \cite{randomized-reversible} present an obfuscation technique based on the addition of a random quantum circuit composed of gates of the original circuit, to the original circuit before compilation. An inverse of the random circuit is separately compiled and concatenated after compilation in order to restore functionality. While their approach uses quantum reversibility, ours differs by classically correcting corrupted outputs without quantum reversals. Their approach can be vulnerable to information leakage if the two circuits are given to the same or related compilers, and compiler gate library variations can hinder physical-to-virtual (p2v) qubit map continuity, which needs careful insertion. If a random circuit is inserted in the middle of a circuit, barriers would be needed for post-compilation detection, restoring the security-optimization tradeoff. While their enhanced method enhances random circuit insertion for quantum circuits, our approach uniquely relies on classical post-processing to deobfuscate without requiring quantum reversals or barriers.
        
        Logic locking methods, a hardware security technique that embeds a key-dependent transformation to prevent unauthorized use or reverse engineering, have also been explored for quantum circuit obfuscation \cite{logic-locking}. The work in \cite{qll} proposes the addition of controlled gates or the substitution of existing gates with their controlled counterparts, using ancilla qubits as controls in the $\ket{0}$ (idle gate) or $\ket{1}$ (active gate) state, where the ancilla configuration encodes a binary key. An enhancement is presented in \cite{e-loc}, which combines multiple key bits into a single ancilla qubit using a technique called “H-masking,” wherein Hadamard gates are inserted during encryption and selectively replaced with Pauli-X gates during decryption to conceal the key. While effective, these methods increase circuit complexity and may exacerbate crosstalk errors \cite{cross-talk} in noisy NISQ environments, potentially making them impractical under current hardware constraints.
        
        Rehman, Langford, and Liu \cite{opaque} propose an obfuscation method that adjusts phase gate angles. Their approach segments the circuit into layers of phase and non-phase gates, sets rotation angles in the phase layers using a binary key, and inserts a dummy phase gate layer—with barriers between layers—to disrupt circuit functionality. However, the method requires the presence of phase gates, limiting its applicability for circuits without them. Furthermore, the mandatory use of barriers hinders circuit optimization and may leave the dummy layer exposed to attackers. In contrast, our approach avoids barriers and phase-specific constraints, offering broader applicability and enabling straightforward deobfuscation via classical correction.
    
    \section{Metrics}\label{sec:metrics}
        \subsection{Total Variation Distance}
            Total Variation Distance (TVD) is a statistical measure of the difference between two probability distributions\cite{tvd}. TVD is used here as a means of quantifying the degradation achieved by the addition of gates to the circuit. It is computed as the sum of absolute differences between the measurement outcomes of original and obfuscated circuits, normalized by the number of shots, as shown in the equation below. 
            
            \begin{equation}
                \text{TVD} = \frac{\sum\limits_{i}|x_{i, obfus} - x_{i, orig}|}{2 \times \text{Number of shots}}
            \end{equation}
            where $x_{i, orig}$ and $x_{i, obfus}$ refer to the count of the $i^{th}$ measurement output of the circuits, before and after addition of encryptor gates respectively. A TVD of 0 indicates identical distributions, and 1 represents maximum difference between the distributions. Thus, a higher TVD in our case would indicate that there is significant difference between the outputs of original circuit and encrypted circuit, which is desired. 

        \subsection{Degree of Functional Corruption}
            Degree of Functional Corruption (DFC) is a metric introduced by Das and Ghosh \cite{randomized-reversible} to quantify the degree of obfuscation. It is defined as the difference between the count of correct output of the obfuscated circuit and the highest incorrect output of the obfuscated circuit, normalized over the number of shots, given by the equation below.
            
            \begin{equation}
                \text{DFC} = \frac{\text{Count of correct output} - \text{Highest Incorrect Output}}{\text{Number of shots}}
            \end{equation}

            If a circuit is perfectly obfuscated, every measured output would be incorrect. If the circuit produces same incorrect output over all shots, then normalized DFC would be -1. For an unobfuscated circuit, if the correct output is produced over all the shots, DFC would be 1. Thus, DFC lies in the range $[-1, 1]$, with lower values indicating better obfuscation.

    \section{Proposed Obfuscation Technique}\label{sec:proposed-technique}
        \subsection{Methodology}
           This section describes our obfuscation method for quantum circuits. The technique leverages specific properties of quantum gates and quantum measurement to enable classical decryption of corrupted outputs post-measurement.

            The theoretical foundation of our method is based on the following key properties of quantum systems:
            
            \begin{enumerate}[label=\roman*.]
                \item Every quantum state, upon measurement, collapses into the measured state. This means that once a quantum system is measured, it loses the quantum behavior and collapses into a classical state. \label{method-1}
                
                \item Phase changes do not affect measurement outputs. If a gate is applied to a qubit such that it introduces only a phase change (such as the S-gate, T-gate, Z-gate, and so on), upon measurement, it would be as though the gate was never applied. Consider a qubit,
                
                \begin{equation}
                    \ket{\psi} = \alpha \ket{0} + \beta \ket{1}
                \end{equation}
                
                After applying an arbitrary phase gate, which introduces a phase of angle $\theta$, the state of qubit changes as
                
                \begin{equation}
                    \ket{\psi'} = \alpha \ket{0} + e^{i\theta}\beta \ket{1}
                \end{equation}
    
                The probability amplitudes of measuring $\ket{\psi'}$ as a 0 or 1 remain exactly same as that of $\ket{\psi}$, as $|e^{i\theta}| = 1$. \label{method-2}
                
                \item A gate that changes the state of a qubit, from $\ket{0}$ to $\ket{1}$ and vice-versa, along with their controlled versions, introduce a definite bit flip to the measured output (if the gate introduced is not a controlled-gate), or introduce bit flip based on some other bit (if the gate is a controlled gate). The effect introduced by such gates to the measurement output can be reversed classically, by application of appropriate classical gates. A few examples are noted in Fig. \ref{fig:equivalent-gates}. \label{method-3}
    
                \begin{figure}[htbp!]
                    \centering
                    \includegraphics[width=1\linewidth]{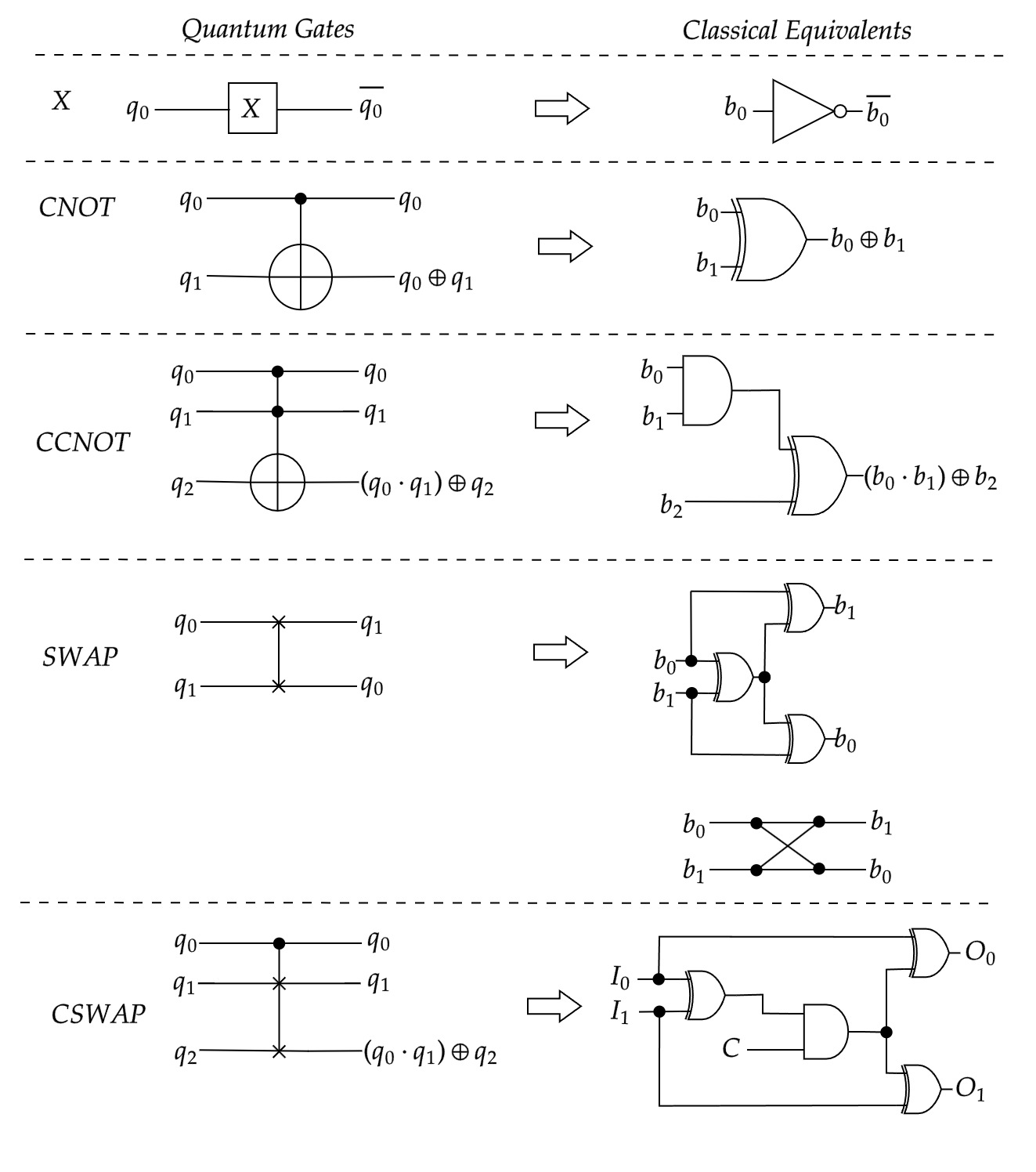}
                    \caption{Some quantum gates with their classical equivalents}
                    \label{fig:equivalent-gates}
                \end{figure}

                \item A gate that introduces both phase change and state change (such as Pauli Y-gate) will impact the measurement output as a sort of combination of rules (\ref{method-2}) and (\ref{method-3}) above.

            \end{enumerate}

            These properties form the basis for the following stages of our obfuscation technique:
    
            \subsubsection{Indexing Gates}

                The first step for applying our obfuscation scheme is to identify the gates that are to be added to the circuit. These gates can be state-changing gates (such as Pauli-X), phase-changing gates (such as Pauli-Z) or a combination of both (such as Pauli-Y), as well as their controlled versions. Hadamard gates and their controlled variants are deliberately excluded from our encryption process, as they introduce superposition and alter the measurement outcome space. Since our technique relies on classical post-processing to decrypt results, the non-deterministic nature of superposition makes it infeasible to reverse such transformations classically.
    

                Once the set of gates is chosen, each of the individual gates within the set must be assigned a unique index. This index serves as a compact identifier in the key and ensures unambiguous decoding during decryption. The indexing scheme is user-defined, but it should be consistent and fixed for a given execution. Controlled versions of gates do not require their own separate indexing scheme; any gate, controlled or uncontrolled, can be assigned any unique index as long as it is used consistently throughout the encryption process. This abstraction maintains the key light-weighted and structured even in the case of multiple gates.

            \begin{figure*}[t]
                \centering
                \includegraphics[width=1\linewidth]{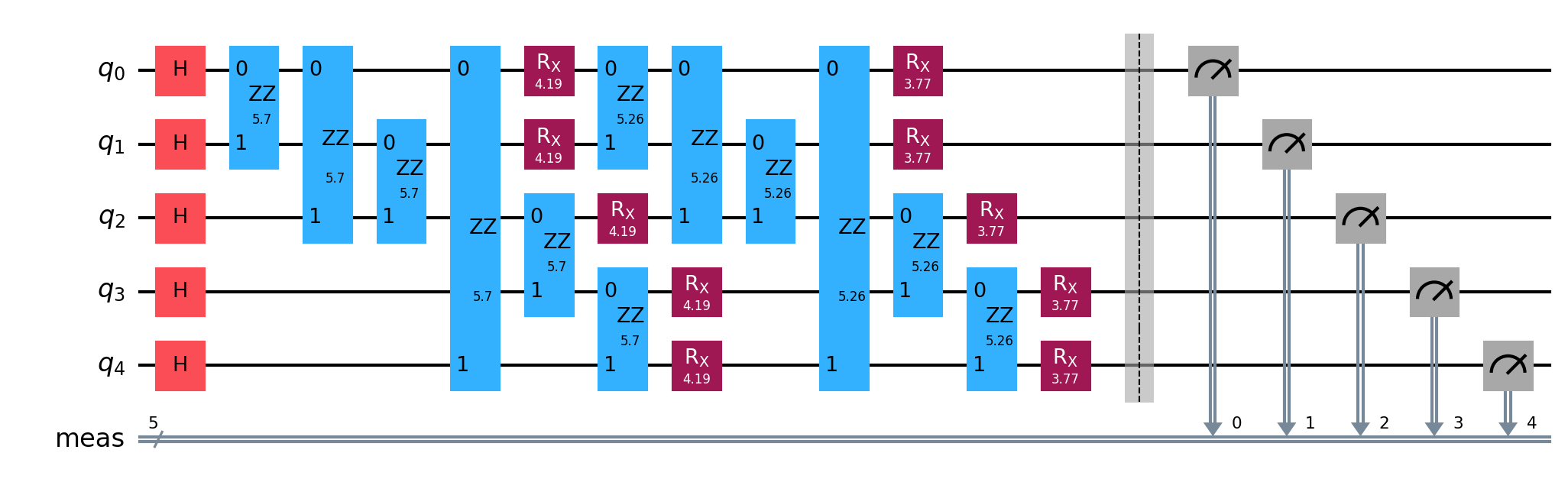}
                \caption{QAOA Circuit}
                \label{fig:qaoa-original}
            \end{figure*}

            \FloatBarrier
    
            \subsubsection{Adding Gates and Key Generation}
                
                After the gate set has been finalized and indexed, the next step is to append the selected gates to the end of the original quantum circuit. The user determines which gate to add and to which qubit(s) it should be applied. These additions serve as the encryption layer of the obfuscation process.
                
                To generate the key, the index of each gate, along with the corresponding qubit(s) it is applied to, is recorded. The format of the key includes delimiters or markers to separate gate indices from qubit references, as well as to distinguish between multiple gate-qubit entries. Importantly, the decryption process applies corrections in the reverse order of insertion. Therefore, the most recently added gate-qubit pair is placed at the beginning of the key, maintaining a stack-like (LIFO) \cite{stack} structure.
                
                Although the selection of gates and their placement is user-defined, this process can also be randomized to enhance the security of the obfuscation. For example, gates can be randomly selected from the valid set and assigned to qubits chosen according to a pseudorandom function, provided that the user retains or stores the corresponding key structure.

            \subsubsection{Running on a Quantum Computer}
                After the obfuscation step has been performed, the obfuscated quantum circuit is then compiled and executed on a quantum computer. In practice, this may involve sending the circuit to a third-party compiler or a quantum service provider. The compiled circuit is then executed for a certain number of shots, and the measurement results are fed back to the user.
                
                Since the inserted gates intentionally manipulate the measuring outcome, the outputs produced by the quantum hardware will not represent the real output of the original circuit. However, the encryption is designed such that the corrupted outputs can be rectified classically using the secret key generated during encryption. The next step is applying this key to restore the correct output, as detailed below.
                
            \subsubsection{Correcting outputs post measurement}
                Once the incorrect measurement output is obtained, the key---which is only available to the user---can be used to extract the correct output. The markers set in the key help in separating the gate-qubit sets, and subsequently the gate and qubit(s) from each set. Once the gate and qubit(s) are identified, the output is corrected classically, by applying appropriate classical operations as shown in fig. \ref{fig:equivalent-gates}. Gates that introduce only a phase change to the qubit can be ignored while correcting measurement output, as these gates do not affect measurement value. Classical operations for gates other than those shown in Fig. \ref{fig:equivalent-gates} can also be derived in a similar manner.
            

            
        \subsection{Case Study: QAOA circuit}
            The obfuscation technique has been demonstrated for the case of the Quantum Approximate Optimization Algorithm (QAOA) circuit. QAOA is a hybrid quantum-classical algorithm used for combinatorial optimization problems \cite{qaoa}. We have considered the implementation of QAOA for the MaxCut problem, demonstrated in IBM Qiskit Learning \cite{ibm-qaoa}. 
    
            The QAOA circuit used is shown in Fig. \ref{fig:qaoa-original}. The circuit shown is the decomposed version of QAOA Ansatz taken in IBM Qiskit Learning.
            

            

            
\begin{figure}[htbp!]
                \centering
                \includegraphics[width=1\linewidth]{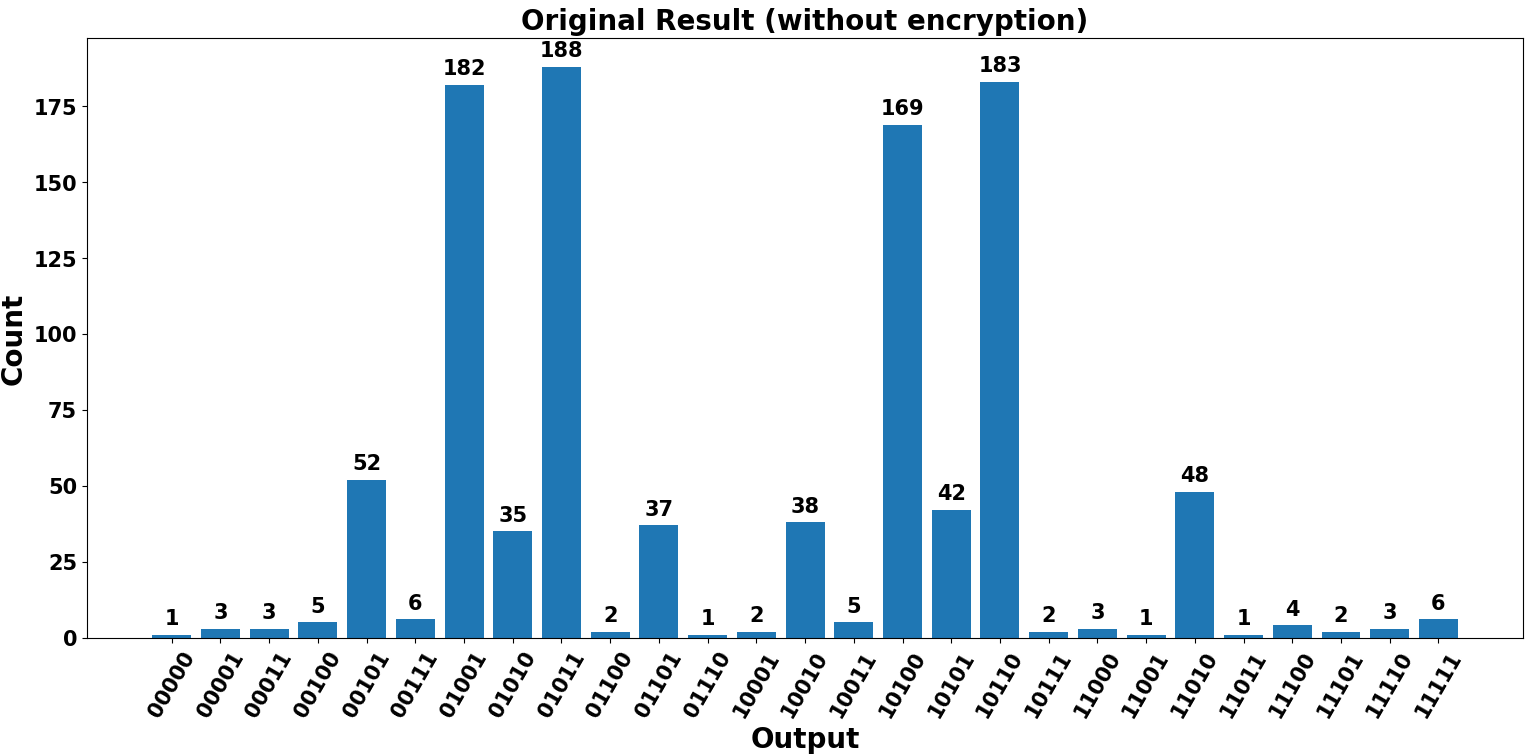}
                \caption{Histogram showing measurement count of the QAOA circuit}
                \label{fig:qaoa-result}
            \end{figure}

            Upon measurement of the final output, we get the histogram as shown in Fig. \ref{fig:qaoa-result}. Note that 01001 and 01011, along with their complements 10110 and 10100 respectively, have the highest peaks, indicating that the vertices 0, 2, 3 can be grouped together, and vertices 1, 4 form the other group.
    
            \FloatBarrier

            \begin{figure*}[t]
                \centering
                \includegraphics[width=1\linewidth]{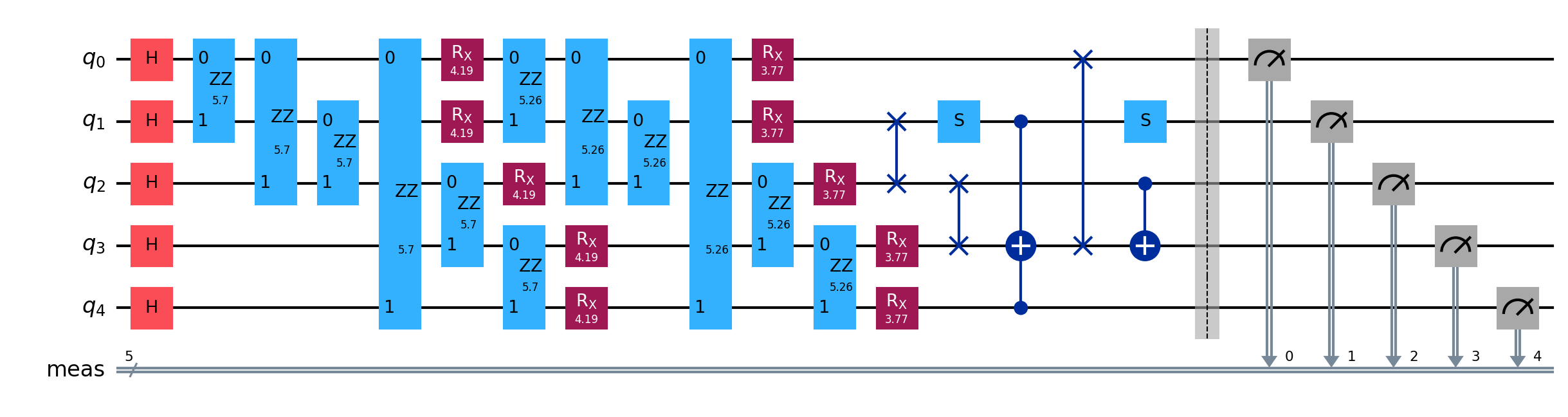}
                \caption{QAOA circuit with newly added gates}
                \label{fig:qaoa-encrypted}
            \end{figure*}

            \FloatBarrier
    
            Now, some extra gates are added to corrupt the measurement outputs. The circuit along with the added gates is shown in Fig. \ref{fig:qaoa-encrypted}.
    

    
\begin{figure}[htbp!]
                \centering
                \includegraphics[width=1\linewidth]{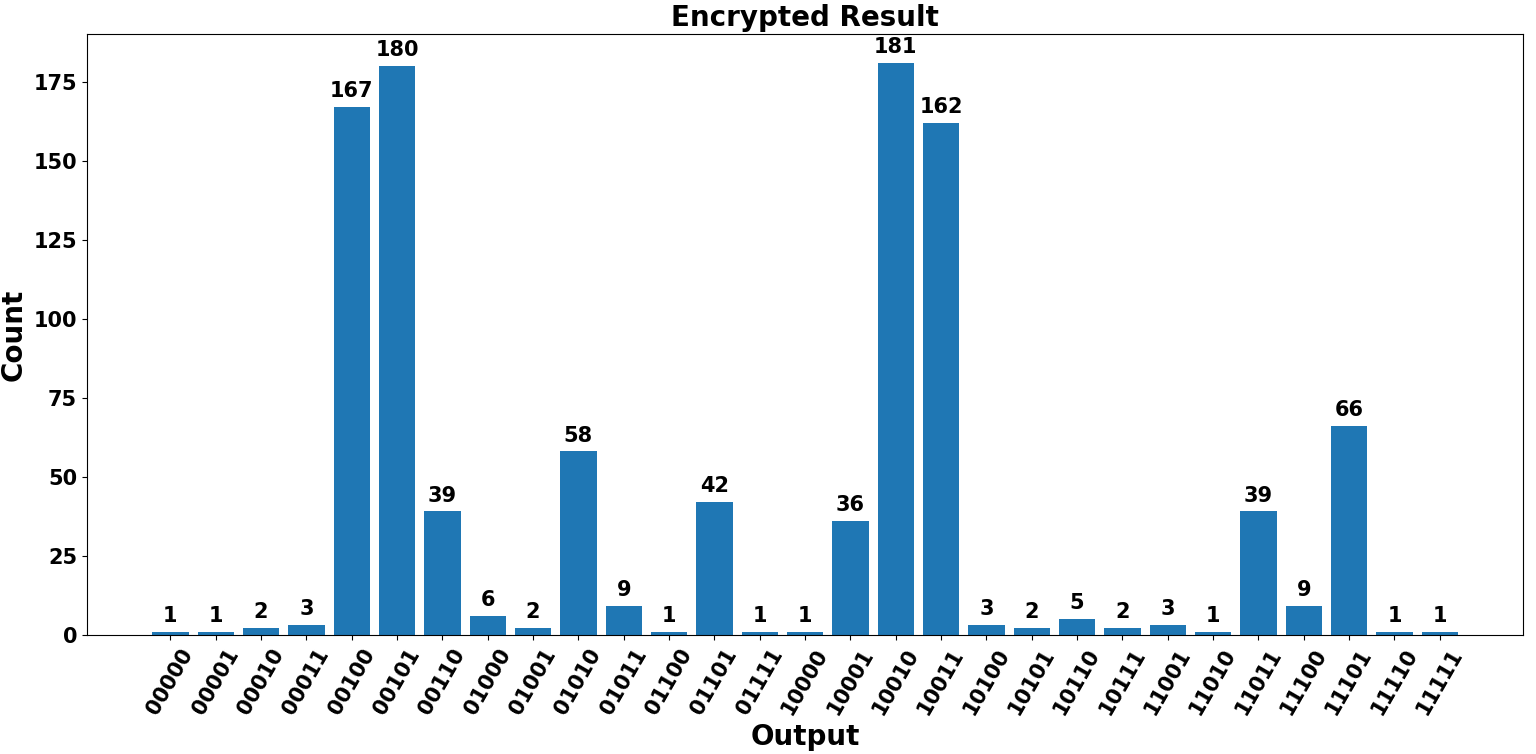}
                \caption{Measurement results after addition of extra gates}
                \label{fig:qaoa-encrypted-result}
            \end{figure}

            The addition of the extra gates has totally corrupted the measurement outputs, which is evident in the histogram shown in Fig. \ref{fig:qaoa-encrypted-result}. It can be seen that peaks in the histogram, now correspond to different bitstrings compared to original results, indicating that the output has been degraded.
            
            \FloatBarrier
    
            The results are now to be corrected using the gate-index mapping generated prior, and the key obtained after the addition of extra gates. For this case, the following information is available:
            
            \begin{itemize}
                \raggedright
                \item The key obtained: \textit{``2\#2$\lvert$1@5\#1@2\#2$\lvert$3@3\#1$\lvert$4$\lvert$3@2\#3$\lvert$0@5\#1@1\#2$\lvert$3''}, with ``@'', ``$\#$'' and ``$|$'' being the demarcators --- ``@'' separating one gate-qubit(s) set from another, ``$\#$'' separating gate from qubit(s), and ``$|$'' separating the qubits.
                \item From the key, the index-qubit(s) set is obtained to be the following --- (2, 2, 1), (5, 1), (2, 2, 3), (3, 1, 4, 3), (2, 3, 0), (5, 1), (1, 2, 3). This is the order that will be used while correcting the output later.
                \item The gate indexing is the following --- \{0: ``X'', 1: ``CNOT'', 2: ``SWAP'', 3: ``CCNOT'', 4: ``CSWAP'', 5: ``S''\}.
            \end{itemize}
            
            The above information can now be used to correct the measured output. For each group in the index-qubit(s) set obtained above, apply the classical equivalent of the corresponding quantum gate, as shown in Fig. \ref{fig:equivalent-gates}, to the bits indicated by the qubit number. If the quantum gate changes only the phase of a qubit, an identity operation can be considered as its classical equivalent, and the bits remain unaffected. The steps for decryption in this particular case is written below:
            
            \begin{enumerate}
                \item (2, 2, 1) --- Swap the bits 1 and 2.
                \item (5, 1) --- Identity operation as the ``S'' gate is a quantum phase gate.
                \item (2, 2, 3) --- Swap the bits 2 and 3.
                \item (3, 1, 4, 3) --- Flip the bit 3 if bits 1 and 4 are set, else do nothing.
                \item (2, 3, 0) --- Swap the bits 3 and 0.
                \item (5, 1) --- Identity operation as the ``S'' gate is a quantum phase gate.
                \item (1, 2, 3) --- Flip the bit 3 if bit 2 is set, else do nothing.
            \end{enumerate}
    
            The corrected output after performing the above operations is shown in the Fig. \ref{fig:qaoa-decrypted} below. The peaks now correspond to the same values that were obtained prior to the addition of gates, indicating the incorrect results were correctly decoded to the actual measurement results. An anonymized GitHub link for the project is available at \cite{anonymous}.
    
            \begin{figure}[htbp!]
                \includegraphics[width=1\linewidth]{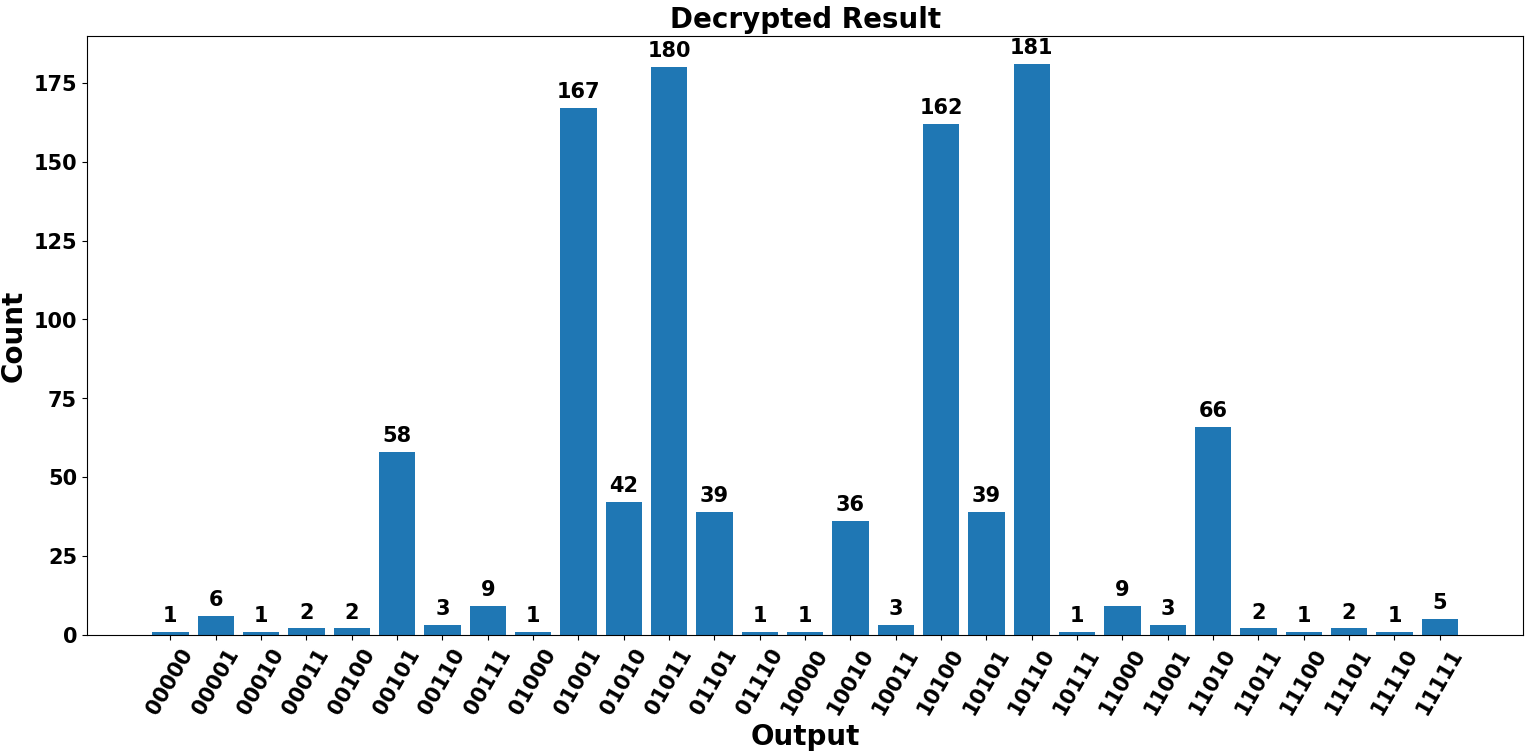}
                \caption{Corrected measurement results}
                \label{fig:qaoa-decrypted}
            \end{figure}

            \FloatBarrier

    \section{Evaluation} \label{sec:evaluation}
        This section presents the experimental evaluation of our proposed obfuscation scheme. All simulations were carried out using IBM’s Qiskit framework \cite{qiskit} on a noise-free AerSimulator backend. The evaluation is divided into two parts: the implementation setup used to apply encryption, and the performance results based on statistical metrics across various quantum algorithms.

        \subsection{Implementation Setup}
            All simulations were performed locally on a laptop with an AMD Ryzen 5 5500U processor (2.10 GHz), 8 GB RAM, and running Windows 11 Home. Five standard quantum algorithms were selected for evaluation: Shor’s algorithm \cite{shor}, the Quantum Approximate Optimization Algorithm (QAOA) \cite{qaoa}, the Bernstein-Vazirani (BV) algorithm \cite{bv}, Grover’s algorithm \cite{grover}, and the Harrow-Hassidim-Lloyd (HHL) algorithm \cite{hhl}. These cover a broad range of application fields—ranging from optimization, factorization, and linear systems—to enable us to gauge the overall applicability of the proposed obfuscation method.

            The quantum circuits varied in size, ranging from 4, 5, or 10 qubits depending on the algorithm. Each algorithm was executed 100 times, and each execution was done for 1,024 shots under identical, noise-free conditions. This ensured that all differences in output distribution could be attributed solely to the encryption process rather than noise or hardware-induced variability.
            
            The encryptor gates belonged to a gate pool shared across all algorithms. Gates were chosen at random from this pool and applied to randomly chosen qubits in each experiment. In spite of the fact that this gate pool was shared among all algorithms, the method still realized good obfuscation effects. This suggests that even generic encryption gate sets can be broadly effective. Future work may investigate the benefits of algorithm-specific gate sets to further optimize obfuscation performance.

        \subsection{Evaluation Metrics}
\begin{figure}[htbp!]
                \centering
                \includegraphics[width=1\linewidth]{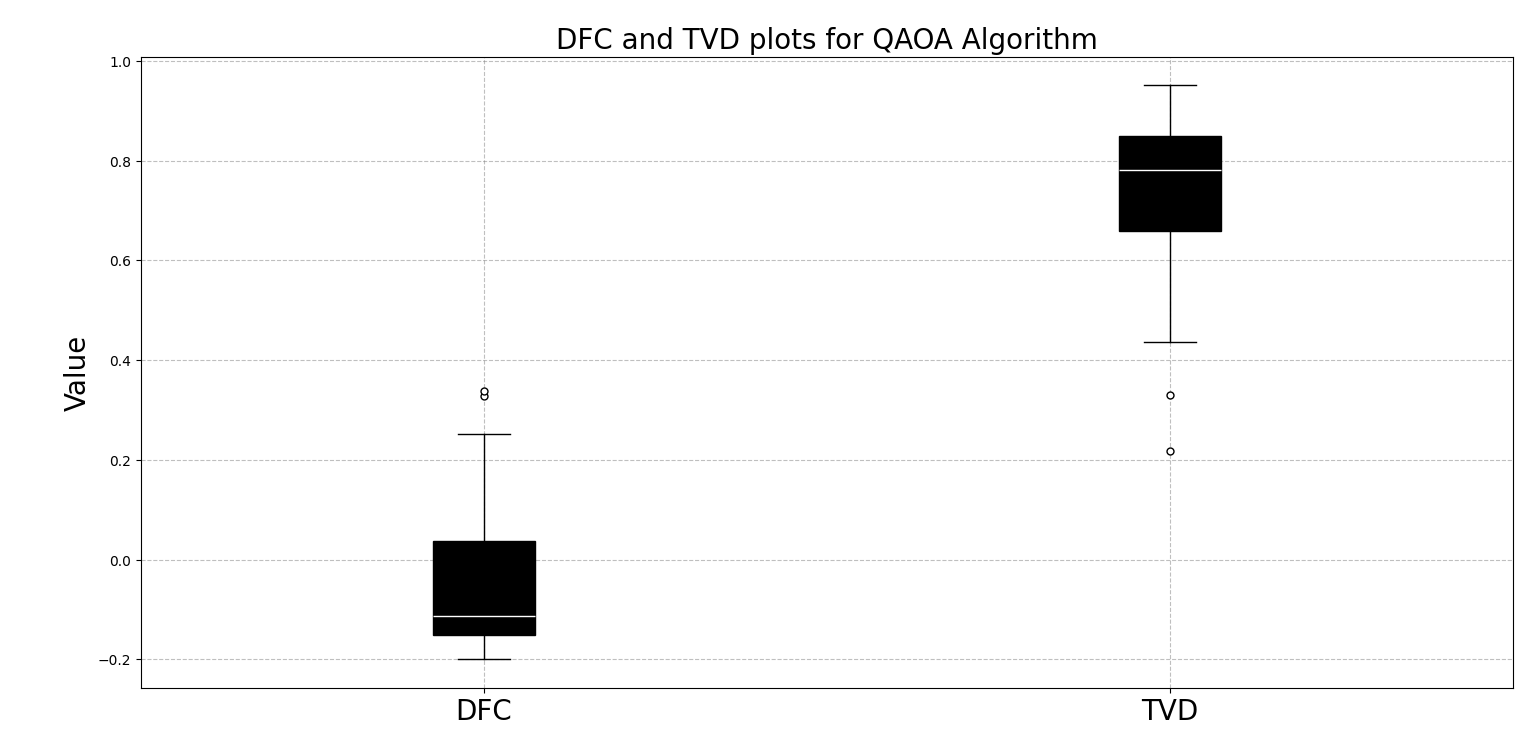}
                \caption{Distribution of TVD and DFC for QAOA implementation. QAOA circuit was obfuscated 100 times by addition of random gates each time, and the TVD and DFC values were computed each time.}
                \label{fig:boxplot-qaoa}
            \end{figure}

\begin{figure}[htbp!]
                \centering
                \includegraphics[width=1\linewidth]{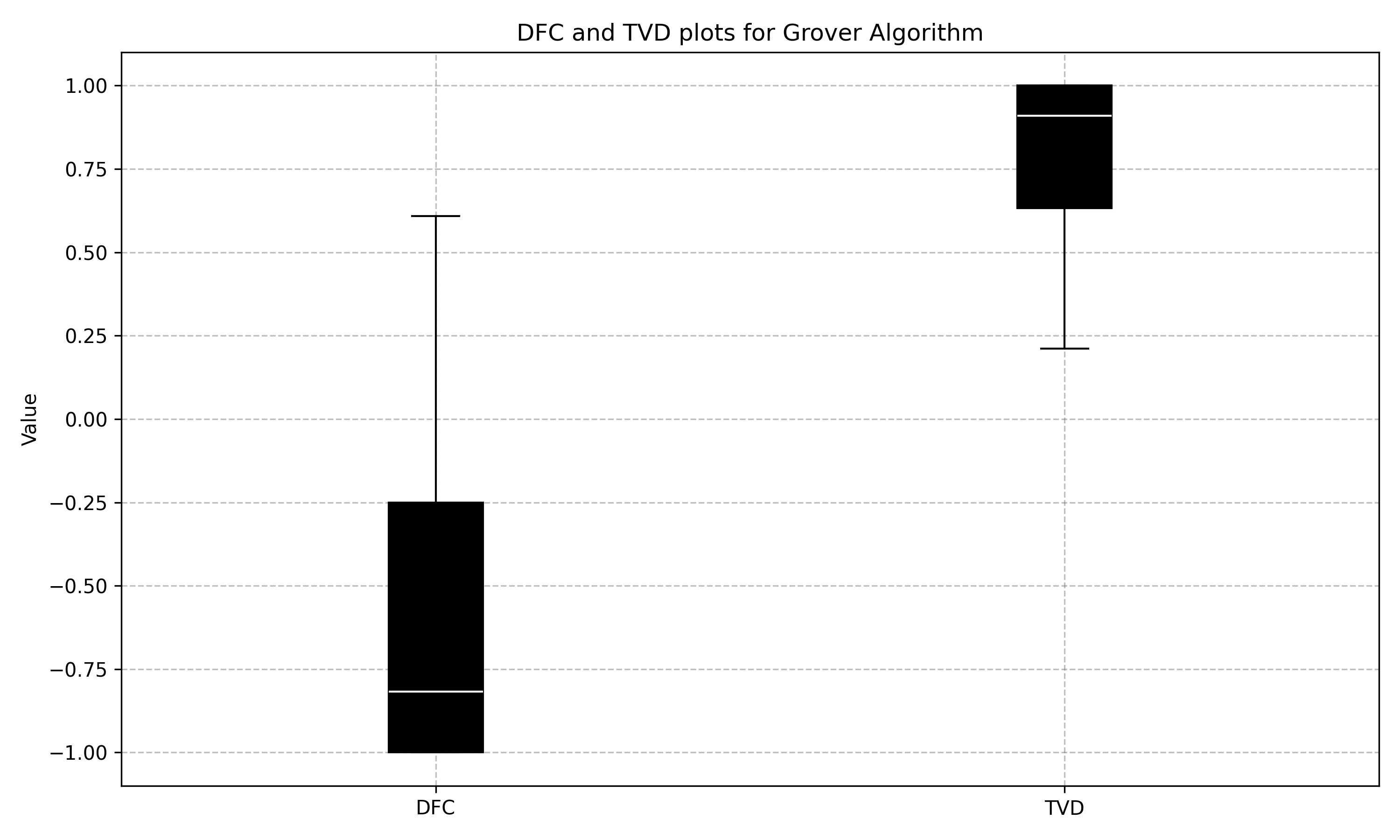}
                \caption{Distribution of TVD and DFC for Grover's algorithm implementation}
                \label{fig:boxplot-grover}
            \end{figure}

\begin{figure}[htbp!]
                \centering
                \includegraphics[width=1\linewidth]{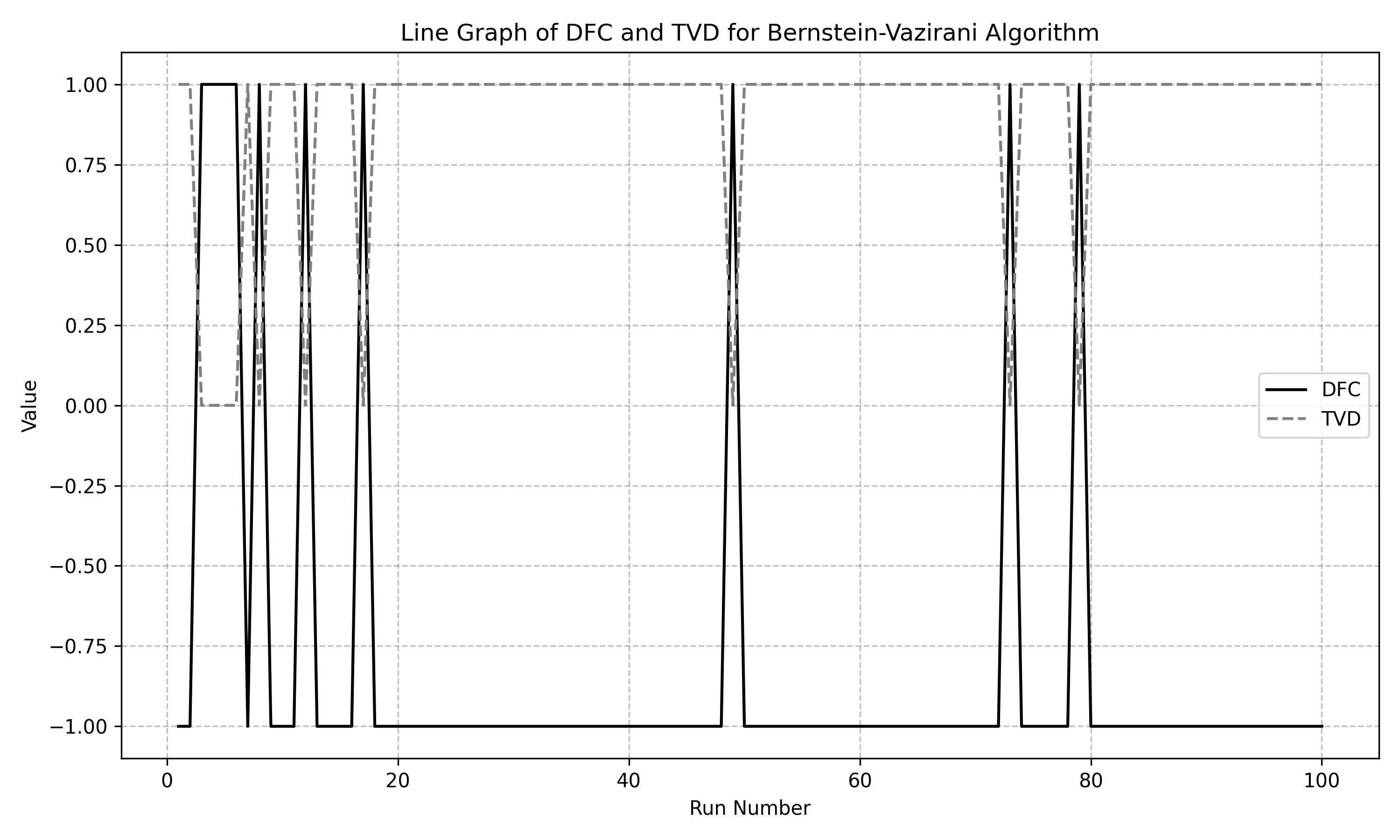}
                \caption{Distribution of TVD and DFC for Bernstein-Vazirani algorithm implementation}
                \label{fig:boxplot-bv}
            \end{figure}

\begin{figure}[htbp!]
                \centering
                \includegraphics[width=1\linewidth]{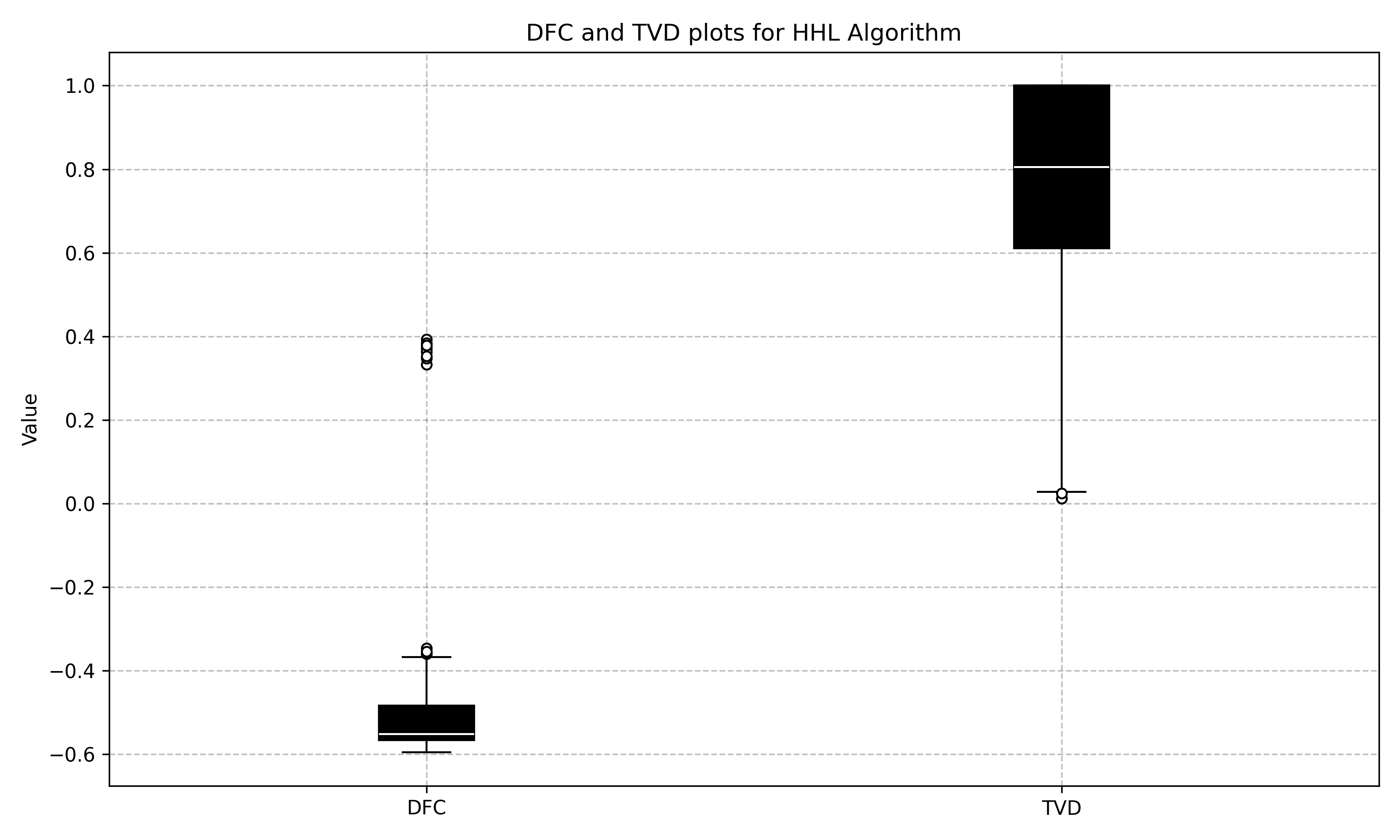}
                \caption{Distribution of TVD and DFC for HHL algorithm implementation}
                \label{fig:boxplot-hhl}
            \end{figure}

\begin{figure}[htbp!]
                \centering
                \includegraphics[width=1\linewidth]{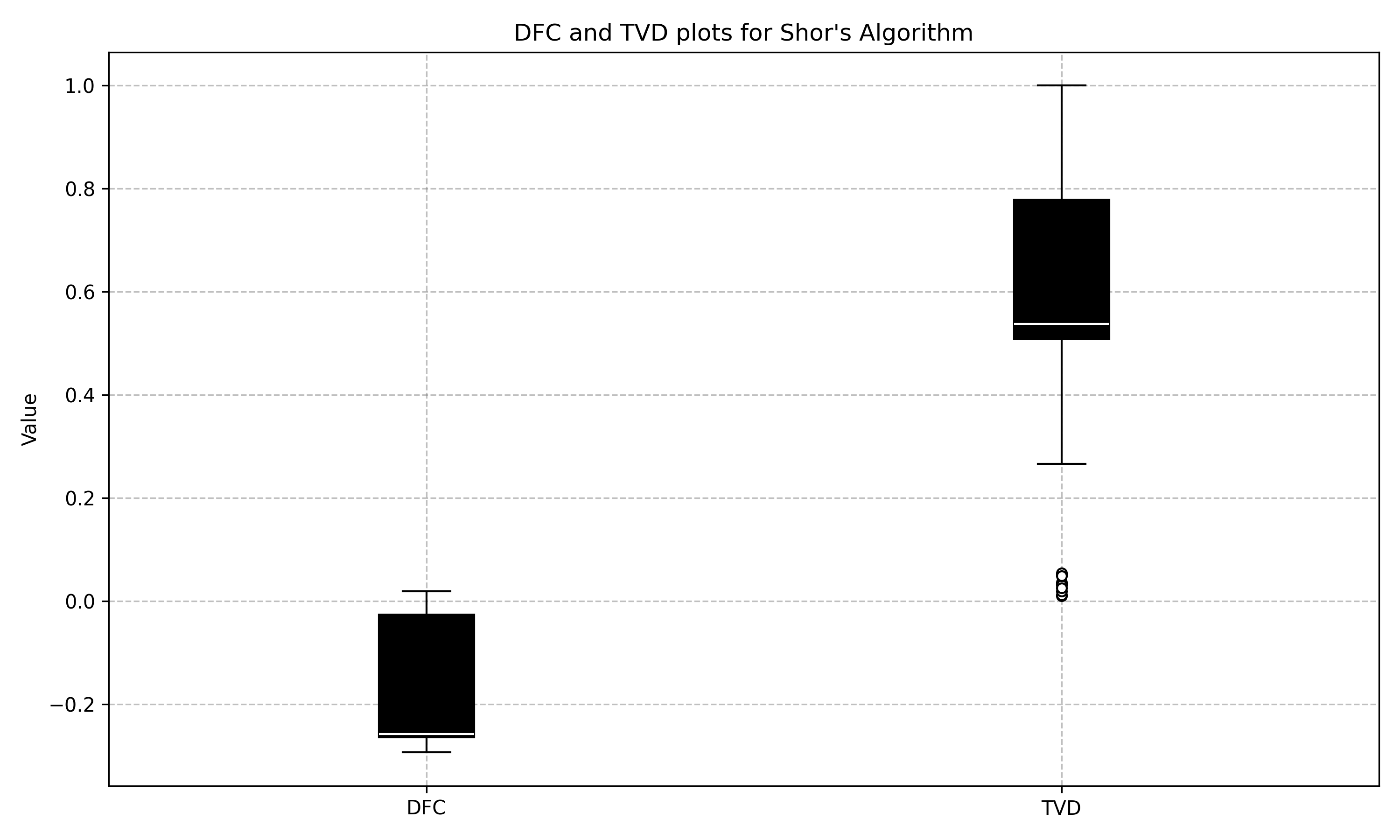}
                \caption{Distribution of TVD and DFC for Shor's algorithm implementation}
                \label{fig:boxplot-shor}
            \end{figure}

            The obfuscation quality achieved by our technique was quantitatively evaluated using Total Variation Distance (TVD) and Degree of Functional Corruption (DFC) metrics. These metrics were computed for every execution of the five algorithms,  results of which are summarized in the Table \ref{tab:metrics} below. Distribution of metrics across the various executions can be observed through the graphs shown in Figs. \ref{fig:boxplot-qaoa}--\ref{fig:boxplot-shor}.

            \renewcommand{\arraystretch}{1.3}
            \begin{table}[h!]
                \centering
                \caption{Median TVD and DFC across the quantum algorithms}
                \begin{tabularx}{1\linewidth}{|Y|Y|Y|}
                    \hline
                    \textbf{Algorithm Tested} & \textbf{Median TVD} & \textbf{Median DFC} \\
                    \hline
                    Shor's & 0.5381 & -0.2583 \\
                    \hline 
                    QAOA & 0.7817 & -0.1123 \\
                    \hline 
                    BV & 1 & -1 \\
                    \hline 
                    Grover's & 0.9097 & -0.8169 \\
                    \hline 
                    HHL & 0.8047 & -0.5522 \\
                    \hline
                \end{tabularx}
                \label{tab:metrics}
            \end{table}

            \FloatBarrier

            It can be observed that the median TVD value is consistently above $0.5$ for all the tested algorithms, indicating that there is a significant change in the output distribution caused by the encryption. This statistical divergence indicates that the obfuscated circuits produce outputs that are distinct from their original counterparts, thereby effectively concealing the underlying functionality.
            
            Similarly, the median DFC scores are negative for all five algorithms, confirming that the encryption scheme corrupts functional behavior. This means that, aside from statistical divergence, the obfuscation also disrupts the functional correctness of the circuits such that the outputs become unreliable in the absence of the decryption key. While Bernstein-Vazirani algorithm and Grover’s algorithm demonstrate strong obfuscation, with DFC values approaching $-1$, QAOA and Shor's algorithm exhibit more moderate DFC values. This variation could be due to the differences in structure of the algorithms, or more likely, due to the uniform gate pool applied across all circuits. Future work could explore algorithm-specific encryption strategies to further enhance functional disruption where needed.
            
            Together, these results demonstrate that the proposed obfuscation scheme achieves both strong statistical distortion and reliable functional corruption across a range of quantum algorithms.

    \section{Limitations and Future Works}\label{sec:limitations}
        While our proposed quantum obfuscation technique yields promising results in protecting hidden quantum circuits, it is restricted by some limitations worthy of greater investigation. First, the current framework adopts random injection of quantum gates (excluding Hadamard gates) to compromise measurement outputs. While randomness can readily conceal the behavior of the circuit, under other circumstances it does not always completely optimize corruption. Selecting gates strategically, in a gate-specific and structure-specific way as opposed to that of the initial circuit, may enhance the functional corruption extent as measured by counts like Total Variation Distance (TVD) and Degree of Functional Corruption (DFC). Research into heuristic or algorithmic selection of gates can lead to safer obfuscation. Second, the current framework is designed for circuits with measurements performed in a single basis per execution, even if that basis is transformed (e.g., measuring in the X-basis using Hadamard gates). However, the technique faces limitations when applied to quantum algorithms that require simultaneous or repeated measurements in multiple bases—such as in variational quantum algorithms that involve Hamiltonians composed of terms across different Pauli bases (e.g., $XX + ZY + YI$). In such scenarios, a single obfuscation layer may not support all measurement contexts without affecting the correctness or requiring significantly more complex post-processing.

        We plan to overcome these disadvantages in future work by developing procedures for the intelligent selection of encryptor gates to maximize corruption at the expense of not compromising classical post-processing simplicity. Based on observation of the connectivity and gate composition of the original circuit, we will devise procedures which artfully insert gates with a view to achieving maximum obfuscation performance. Additionally, we intend to extend the current correction mechanism to support circuits that require measurements in multiple bases within a single execution. This will involve adapting the standard correction procedure to support measurements in other bases, perhaps through the inclusion of basis transformations or generalized correction circuits. These advancements will broaden the applicability of our method to more quantum algorithms and make it more valuable for quantum intellectual property protection in untrusted compilation environments. Besides, we plan to experiment with the scheme in practical NISQ hardware conditions, incorporating noise models to assess its performance and robustness in real-world settings.


    \section{Conclusion}\label{sec:conclusion}
        In this work, we introduced a novel quantum circuit obfuscation technique designed to defend commercial quantum algorithms within the NISQ era, where third-party compiler trust can be an intellectual property vulnerability. By introducing quantum gates, with the exception of Hadamard gates, into the circuit before compilation in a carefully constructed form, our approach is able to corrupt measurement outcomes, which are subsequently corrected by a simple classical post-processing operation. Conversely, with other methods depending on complicated quantum reversals, barriers, or physical-to-virtual qubit mapping concerns, our method reduces deobfuscation complexity and enhances practicality and security. Our technique was evaluated across multiple quantum algorithms and consistently demonstrated high obfuscation strength, both in terms of statistical divergence and functional corruption. These results confirm that the method is broadly applicable across a range of circuit structures, offering a lightweight yet effective layer of security. Overall, this work takes a significant step toward securing quantum computations in untrusted environments, offering a practical and scalable obfuscation solution for real-world quantum applications.

\end{document}